\documentclass[twocolumn,showpacs,aps,prl,superscriptaddress]{revtex4}
\usepackage[dvips]{graphicx}
\usepackage{dcolumn}
\usepackage{epsfig}
\usepackage{amsmath}
\RequirePackage{xspace}

\usepackage{relsize}
\def\babar{\mbox{\slshape B\kern-0.1em{\smaller A}\kern-0.1em
    B\kern-0.1em{\smaller A\kern-0.2em R}}}
\def\CP                {\ensuremath{C\!P}\xspace}
\def\ra                 {\ensuremath{\rightarrow}\xspace}
\def\to                 {\ensuremath{\rightarrow}\xspace}
\def\calA{{\ensuremath{\cal A}}\xspace}

\def\Bbar    {\kern 0.18em\overline{\kern -0.18em B}{}\xspace}

\def\BB      {\ensuremath{B\Bbar}\xspace} 
\def\epem       {\ensuremath{e^+e^-}\xspace}

\newcommand{\mev}{\ensuremath{\mathrm{\,Me\kern -0.1em V}}\xspace}

\def\Bz      {\ensuremath{B^0}\xspace}

\newcommand{\etapr}{\ensuremath{\eta^{\prime}}\xspace}
\newcommand{\gev}{\ensuremath{\mathrm{\,Ge\kern -0.1em V}}\xspace}
\def\Bzb     {\ensuremath{\Bbar^0}\xspace}

\newcommand{\kspzpz}{\mbox{$\KS \rightarrow\pi^0\pi^0$}}
\newcommand{\kspppm}{\mbox{$\KS\rightarrow\pi^+\pi^-$}}
\newcommand{\fetapreppthrpi}{\ensuremath{\etapr_{\eta(3\pi)\pi\pi}}}

\newcommand{\fetapreppgg}{\ensuremath{\etapr_{\eta(\gamma\gamma)\pi\pi}}}

\newcommand{\fetapreppggkz}{\ensuremath{\etapr_{\eta(\gamma\gamma)\pi\pi} K^{0} }}
\newcommand{\fetapreppggk}{\ensuremath{\etapr_{\eta(\gamma\gamma)\pi\pi} K^{+} }}

\newcommand{\fetapreppthrpikz}{\ensuremath{\etapr_{\eta(3\pi)\pi\pi}K^{0}}}

\def\Bz      {\ensuremath{B^0}\xspace}
\def\Bzb     {\ensuremath{\Bbar^0}\xspace}
\def\BzBzb   {\ensuremath{\Bz {\kern -0.16em \Bzb}}\xspace}
\def\Bu      {\ensuremath{B^+}\xspace}
\def\Bub     {\ensuremath{B^-}\xspace}

\def\BpBm    {\ensuremath{\Bu {\kern -0.16em \Bub}}\xspace}
\providecommand{\DE}{\ensuremath{\Delta E}}
\providecommand{\pvec}{{\bf p}}
\providecommand{\half}{\mbox{${1\over2}$}}
\providecommand{\calB}{\mbox{${\cal B}$}}

\providecommand{\etapKzs}{\mbox{$\eta^{\prime} \KS$}}

\providecommand{\kzs}{\ensuremath{\KS}}
\providecommand{\etapKp}{\mbox{$\etapr K^+$}}
\providecommand{\skz}{\mbox{$S$}}
\providecommand{\ckz}{\mbox{$C$}}
\newcommand{\acp}{\ensuremath{\calA_{ch}}}
\providecommand{\UfourS}{\mbox{$\Upsilon(4S)$}}
\newcommand{\BretapKp}{\mbox{$\calB(B^+\ra\eta^\prime K^+)$}}
\newcommand{\BretapKz}{\mbox{$\calB(B^0\ra\eta^\prime K^0)$}}
\providecommand{\BetapK}{\mbox{$B^+ \rightarrow \eta^{\prime} K^+ $}}
\providecommand{\BetapKz}{\mbox{$B^0 \rightarrow \eta^{\prime} K^0$}}
\providecommand{\BetapKzs}{\mbox{$B^0 \rightarrow \eta^{\prime} \KS$}}
\newcommand{\retapKp}{\ensuremath{68.9\pm2.0 \pm3.2}}
\newcommand{\RetapKp}{\ensuremath{(\retapKp)\times 10^{-6}}}
\newcommand{\retapKz}{\ensuremath{67.4\pm 3.3 \pm3.2}}
\newcommand{\RetapKz}{\ensuremath{(\retapKz)\times 10^{-6}}}
\newcommand{\rSetapKz}{\ensuremath{0.30\pm0.14\pm0.02}}
\newcommand{\rCetapKz}{\ensuremath{-0.21\pm0.10\pm0.02}}
\newcommand{\rAetapKp}{\ensuremath{0.033 \pm 0.028\pm 0.005}}
\newcommand{\RAetapKp}{\ensuremath{[-0.012, 0.078]}}
\newcommand{\etal}{{\em et al.}}
\providecommand{\goto}{\rightarrow}

\def\deltat{\ensuremath{{\rm \Delta}t}\xspace}
\def\deltaS{\ensuremath{{\rm \Delta}S}\xspace}
\providecommand{\dt}{\deltat}
\def\deltamd{\ensuremath{{\rm \Delta}m_d}\xspace}

\def\stwob{\ensuremath{\sin\! 2 \beta   }\xspace}
\def\Kp    {\ensuremath{K^+}\xspace}
\def\KS    {\ensuremath{K^0_{\scriptscriptstyle S}}\xspace}
\def\g     {\ensuremath{\gamma}\xspace}
\def\gaga  {\ensuremath{\gamma\gamma}\xspace}
\def\mes        {\mbox{$m_{\rm ES}$}\xspace}

\newcommand{\etaprrg} {\ensuremath{\etapr \to \rho^0 \g}\xspace}

\def\piz{\mbox{${\pi^{0}}$}}
\providecommand{\EtapEtaPiPi}{\mbox{$\etapr \rightarrow \eta \pi^+ \pi^-$}}
\providecommand{\EtapRhoPi}{\mbox{$\etapr \rightarrow \rho^0  \gamma$}}
\providecommand{\EtaGG}{\mbox{$\eta \rightarrow \gamma  \gamma$}}
\providecommand{\EtaTrePi}{\mbox{$\eta \rightarrow \pi^+ \pi^- \pi^0$}}
\providecommand{\mgg}{\mbox{$m_{\gamma \gamma}$}}
\providecommand{\KSZZ}{\mbox{$\KS \rightarrow \pi^0 \pi^0$}}
\providecommand{\mpipi}{\mbox{$m_{\pi \pi}$}}
\providecommand{\mpipipi}{\mbox{$m_{\pi \pi \pi}$}}
\providecommand{\metap}{\mbox{$m_{ \etapr }$}}
\def\qqbar{\mbox{$q\bar q\ $}}

\def\Bub     {\ensuremath{B^-}\xspace}
\def\Bu      {\ensuremath{B^+}\xspace}

\providecommand{\tcp}{\mbox{$t_{CP}$}}
\def\BzBzb   {\ensuremath{\Bz {\kern -0.16em \Bzb}}\xspace}
\def\BpBm    {\ensuremath{\Bu {\kern -0.16em \Bub}}\xspace}
\newcommand{\gevc}{\ensuremath{{\mathrm{\,Ge\kern -0.1em V\!/}c}}\xspace}
\newcommand{\mevc}{\ensuremath{{\mathrm{\,Me\kern -0.1em V\!/}c}}\xspace}
\newcommand{\gevcc}{\ensuremath{{\mathrm{\,Ge\kern -0.1em V\!/}c^2}}\xspace}
\newcommand{\mevcc}{\ensuremath{{\mathrm{\,Me\kern -0.1em V\!/}c^2}}\xspace}

\newcommand{\ttag}{\ensuremath{t_{\rm tag}}}

\newcommand{\etagg}{\ensuremath{\eta\ra\gaga}}
\newcommand{\mb}{\mes}

\newcommand{\xf}{\mbox{${\cal F}$}}
\newcommand{\bflav}{\ensuremath{B_{\rm flav}}}
\providecommand{\sigdt}{\ensuremath{\sigma_{\deltat}}}

\def\dbline{\noalign{\vskip 0.15truecm\hrule}\noalign{\vskip 2pt}\noalign{\hrule\vskip 0.15truecm}}

\providecommand{\tbline}{\noalign{\vskip 0.05truecm\hrule\vskip0.05truecm}}

\newcommand{\fetaprg}{\ensuremath{\etapr_{\rho\gamma}}}

\newcommand{\fetaprgKp}{\ensuremath{\etapr_{\rho\gamma} K^+}}

\newcommand{\fetaprgKz}{\ensuremath{\etapr_{\rho\gamma} K^0}}

\newcommand{\fetapKp}{\ensuremath{\etapr K^+}}

\newcommand{\fetapKz}{\ensuremath{\etapr K^0}}

\def\pep2{PEP-II}

\newcommand{\jprlBase}       {Phys.\ Rev.\ Lett.\xspace}
\newcommand{\jprl}      [1]  {\jprlBase\ {\bf #1}}

\newcommand{\jprBase}        {Phys.\ Rev.\xspace}
\newcommand{\jprd}      [1]  {\jprBase\ D~{\bf #1}}

\newcommand{\progtp}    [1]  {{Prog.\ Theor.\ Phys.\ {\bf #1}}}
\newcommand{\jplBase}        {Phys.\ Lett.\xspace}
\newcommand{\plb}       [1]  {\jplBase\ B~{\bf #1}}
\newcommand{\npBase}         {Nucl.\ Phys.\xspace}
\newcommand{\npb}       [1]  {\npBase\ B~{\bf #1}}
\newcommand{\nimBaseA}       {Nucl.\ Instr.\ Meth.\xspace}
\newcommand{\nima}      [1]  {\nimBaseA~A~{\bf #1}}
\def\Kz    {\ensuremath{K^0}\xspace}

\newcommand{\BaBarYear}    {04}
\newcommand{\BaBarNumber}  {050}

\newcommand{\SLACPubNumber} {10906}

 \newcommand{\BaBarType}      {PUB}  

\begin{document}

\preprint{\babar-PUB-\BaBarYear/\BaBarNumber}
\preprint{SLAC-PUB-\SLACPubNumber}

\begin{flushleft}
 \babar-\BaBarType-\BaBarYear/\BaBarNumber \\
 SLAC-PUB-\SLACPubNumber\\
\end{flushleft}

\title{ \large \bf\boldmath Measurements of  Branching Fractions and 
Time-Dependent $CP$-Violating Asymmetries in  $B\to\etapr K$ Decays }

%
\author{B.~Aubert}
\author{R.~Barate}
\author{D.~Boutigny}
\author{F.~Couderc}
\author{Y.~Karyotakis}
\author{J.~P.~Lees}
\author{V.~Poireau}
\author{V.~Tisserand}
\author{A.~Zghiche}
\affiliation{Laboratoire de Physique des Particules, F-74941 Annecy-le-Vieux, France }
\author{E.~Grauges-Pous}
\affiliation{IFAE, Universitat Autonoma de Barcelona, E-08193 Bellaterra, Barcelona, Spain }
% ...
\collaboration{The \babar\ Collaboration}
\noaffiliation

\begin{abstract}
We present measurements of the 
$B\ra\etapr K$  branching fractions; for 
\BetapK\ we measure
 also  the time-integrated charge asymmetry \acp, and for   \BetapKzs\ 
the time dependent \CP-violation parameters \skz\ and \ckz.
The data sample corresponds to 232 million \BB\ pairs 
produced by \epem\ annihilation at the \UfourS.
The results are  $\BretapKp = \RetapKp$, $\BretapKz =\RetapKz$, $\acp =\rAetapKp$,
 $\skz = \rSetapKz$, and $\ckz = \rCetapKz$, where the first error
quoted is statistical and the second systematic.
\end{abstract}


\maketitle

Measurements of time-dependent \CP\  asymmetries in $B^0$ meson decays through
a Cabibbo-Kobayashi-Maskawa (CKM) favored $b \rightarrow c \bar{c} s$ amplitude
\cite{babar} have provided a crucial test of the mechanism of \CP\ violation
in the Standard Model (SM) \cite{SM}.  Such decays to a charmonium state plus 
a $K^0$ meson are dominated by a single weak phase.
Decays of $B^0$ mesons to charmless hadronic final states, such as
$\phi K^0$, $K^+ K^- K^0$, $\etapr K^0$, $\piz K^0$ and $f_0(980) K^0$, 
proceed mostly via a single penguin (loop) amplitude with the same weak
phase \cite{Penguin}, but CKM-suppressed amplitudes and multiple particles in 
the loop introduce other weak phases whose contribution is not
negligible \cite{Gross,Gronau,Gronau2,BN,london}.

For the decay \BetapKz, these additional contributions are expected to
be small, 
so the time-dependent asymmetry measurement for this decay provides an 
approximate measurement of \stwob.  Theoretical bounds for the small
deviation \deltaS\ between the time-dependent \CP-violating parameter measured
in this decay and in the charmonium $K^0$ decays have been calculated with 
an SU(3) analysis \cite {Gross,Gronau}. Such bounds have been improved by 
measurements of \Bz\ decays to a pair of neutral light 
pseudoscalar mesons \cite{Isosca,PRD}.
From these and other  measurements, improved
model-independent bounds have been derived \cite{Gronau2}, with the
conclusion that \deltaS\ is expected to be less than 0.10 (with a theoretical
uncertainty less than $\sim$0.03).
Specific model calculations conclude that \deltaS\ is even smaller \cite{BN}.
A significantly larger \deltaS\ could arise from non-SM 
amplitudes \cite{london}.

The time-dependent \CP-violating asymmetry in the decay \BetapKz\ has been 
measured previously by the \babar ~\cite{Previous} and Belle~\cite{BELLE} 
experiments.  In this Letter we update our previous measurements with an 
improved analysis and a data sample  four times larger .  
We also measure the \BetapKz\ and \BetapK\ branching fractions
\cite{conjugates}, and for \BetapK\ the time-integrated charge asymmetry
$\acp = (\Gamma^--\Gamma^+)/(\Gamma^-+\Gamma^+)$ where
$\Gamma^\pm=\Gamma(B^\pm\ra\etapr K^\pm)$. In the SM \acp\ is expected
to be small; a non-zero value would signal direct \CP\ violation in this
channel. 

The data were collected with the 
\babar\ detector~\cite{BABARNIM} at the PEP-II asymmetric $e^+e^-$ collider 
~\cite{pep}.  An integrated
luminosity of 211~fb$^{-1}$, corresponding to
232 million \BB\ pairs, was recorded at the $\Upsilon (4S)$
resonance (center-of-mass energy $\sqrt{s}=10.58\ \gev$).
Charged particles are detected and their momenta measured by the
combination of a silicon vertex tracker (SVT), consisting of five layers
of double-sided detectors, and a 40-layer central drift chamber,
both operating in the 1.5 T magnetic field of a solenoid.
Charged-particle identification (PID) is provided by the average energy
loss in the tracking devices and by an internally reflecting ring-imaging
Cherenkov detector (DIRC) covering the central region.
Photons and electrons are detected by a CsI(Tl) electromagnetic calorimeter.

From a candidate \BB\ pair we reconstruct a \Bz\  decaying into the flavor 
eigenstate $f= \etapKzs $ ($B_{CP}$).  We also reconstruct the vertex of the other
$B$ meson ($B_{\rm tag}$) and identify its flavor.
The difference $\deltat \equiv \tcp - \ttag$
of the proper decay times $\tcp$ and $\ttag$ of the \CP\ and tag $B$ mesons, 
respectively, is obtained from the measured distance between the $B_{CP}$
and  $B_{\rm tag}$ decay vertices and from the boost ($\beta \gamma =0.56$) of 
the \epem system. The \deltat\ distribution is given by:
\begin{eqnarray}
  F(\dt) &=& 
        \frac{e^{-\left|\deltat\right|/\tau}}{4\tau} [1 \mp\Delta w \pm
                                                   \label{eq:FCPdef}\\
   &&\hspace{-1em}(1-2w)\left( S\sin(\deltamd\deltat) -
C\cos(\deltamd\deltat)\right)].\nonumber
\end{eqnarray}
The upper (lower) sign denotes a decay accompanied by a \Bz (\Bzb) tag,
$\tau$ is the mean $\Bz$ lifetime, $\deltamd$ is the mixing
frequency, and the mistag parameters $w$ and
$\Delta w$ are the average and difference, respectively, of the probabilities
that a true $\Bz$\ is incorrectly tagged as a $\Bzb$\ or vice versa.
The tagging algorithm \cite{s2b} has seven mutually exclusive tagging categories of
differing response purities (including one for untagged events that we
retain for yield determinations).  The measured analyzing power, defined as 
efficiency times $(1-2w)^2$ summed over all categories, is $( 30.5\pm 0.6)\%$,
as determined from a large sample 
of $B$-decays to fully reconstructed flavor eigenstates (\bflav).
The parameter $C$ measures direct \CP\ violation.  If $C=0$, then
$S=\stwob+\deltaS$.

Monte Carlo (MC) simulations \cite{geant} of the signal decay modes, \BB\ 
backgrounds, and 
detector response are used to tailor the event selection criteria.  We 
reconstruct $B$ meson candidates by combining a $K^+$ or a \KS\ with an \etapr\
meson.  
We reconstruct \etapr\ mesons through the decays \EtapRhoPi\ (\fetaprg)
and \EtapEtaPiPi\ with \EtaGG\ (\fetapreppgg) or 
\EtaTrePi\ (\fetapreppthrpi).   
For the $K^+$ track we require an associated DIRC Cherenkov angle 
between $-5$ and $+2$ standard deviations ($\sigma$) from the
expected value for a kaon. 
We select $\KS\to\pi^+\pi^-$ decays by requiring the $\pi^+\pi^-$ 
invariant mass to be within 12 \mev\ of the nominal \Kz\ mass and by requiring 
a flight length with significance $>$3$\sigma$. We  select \kspzpz\ decays 
by requiring that the $\pi^0\pi^0$ invariant mass be within 30 \mev\ of the 
nominal \Kz\ mass. 
Daughter pions from \etapr\ 
decays are required to have PID information inconsistent with proton,
electron and kaon hypotheses.  The photon energy 
$E_{\gamma}$ must be greater than 30 \mev\ for \piz\ candidates, 50
(100) MeV for $\eta$ candidates for the \fetapreppggkz\ (\fetapreppggk) samples, 
and greater than 100 \mev\ for \fetaprg\ candidates. We make the following 
requirements on the invariant mass (in \mev): 490 $<$ \mgg $<$ 600  for 
\etagg,  $120 < \mgg  < 150$  for $\pi^0$  ($100 < \mgg  < 155$ in \KSZZ ), 
$510<\mpipi<1000$ for $\rho^0$, $520<\mpipipi< 570$ for \EtaTrePi,
$945<\metap < 970$ for \fetapreppgg, and $930<\metap< 980$ for \fetaprg.

A $B$ meson candidate is characterized kinematically by the energy-substituted mass
$\mes \equiv  \sqrt{(\half s + \pvec_0\cdot \pvec_B)^2/E_0^2 - \pvec_B^2}$ and the 
energy difference $\DE \equiv E_B^*-\half\sqrt{s}$, where 
$(E_0,\pvec_0)$ and $(E_B,\pvec_B)$ are four-momenta of
the \UfourS\ and the $B$ candidate, respectively,
and the asterisk denotes the \UfourS\ rest frame.
We require $|\DE|\le0.2$ GeV and $5.25\le\mes\le5.29\ \gev$.  

To reject the dominant background from continuum $\epem\ra\qqbar$ events 
($q=u,d,s,c$), we use
the angle $\theta_T$ between the thrust axis of the $B$ candidate and
that of
the rest of the tracks and neutral clusters in the event, calculated in
the  \UfourS\ rest frame.  The distribution of $\cos{\theta_T}$ is
sharply peaked near $\pm1$ for combinations drawn from jet-like $q\bar q$
pairs and is nearly uniform for the isotropic $B$ decays; we require
$|\cos{\theta_T}|<0.9$.
From Monte Carlo simulations of \BzBzb\ and \BpBm\ events, we find evidence 
for a small (1--2\%) \BB\ background contribution for the channels with 
\EtapRhoPi, so we have added a \BB\ component to the fit described below
for those channels.

We use an unbinned, multivariate maximum-likelihood fit to extract
signal yields and \CP-violation parameters. We indicate with $j$ the
species of  
event: signal, \qqbar\ combinatorial background, or \BB\
background. 
We use four discriminating variables: 
\mes, \DE, \deltat, and a Fisher discriminant \xf\ \cite{Fisher}. The Fisher discriminant
combines  four variables: the angles with respect to the beam axis of the $B$ 
momentum and $B$ thrust axis in the \UfourS\ rest frame, and the zeroth and second 
angular moments of the energy flow, excluding the $B$ candidate, about the 
$B$ thrust axis \cite{AngMom}.

For each species $j$ and tagging  category       
$c$, we define a total probability density function (PDF) for event $i$ as
\begin{equation}
{\cal P}_{j,c}^i \equiv  {\cal P}_j ( \mes^i ) \cdot {\cal  P}_j ( \DE^i )
\cdot { \cal P}_j( \xf^i ) \cdot 
{ \cal  P}_j (\deltat^i, \sigma_{\deltat}^i;c)\,,
\end{equation}
where $\sigma_{\deltat}^i$ is the error on \deltat\ for event $i$.  
With $n_{j}$ defined to be the number of events of the species $j$
and $f_{j,c}$ the fraction of events of species $j$ for each category $c$,
we write the extended likelihood function for all events belonging to category $c$ as
\begin{eqnarray}
{\cal L}_c &=& \exp{\Big(-\sum_{j} n_{j,c}\Big)}
           \prod_i^{N_c} (n_{\rm sig}f_{{\rm sig},c}{\cal P}_{{\rm sig},c}^{i}\\
                  &&\hspace{-1em}+n_{q\bar{q}} f_{q\bar{q},c}{\cal P}_{q\bar{q}}^{i}
                   +n_{B\bar{B}}f_{B\bar{B},c}{\cal P}_{B\bar{B}}^{i}),\nonumber
\end{eqnarray}
where $n_{j,c}$ is the yield of events of species $j$ found by the fitter 
  in category $c$ and $N_c$ the number of events of category c in the sample. 
We fix both $f_{{\rm sig},c}$ and 
$f_{B\bar{B},c}$ to $f_{\bflav,c}$, the values measured with the large
\bflav\ sample  \cite{Resol}. 
The total likelihood function ${\cal L}_d$ for decay mode $d$ is given as the
product over the seven tagging categories.  Finally, when combining
decay modes we form the grand likelihood ${\cal L}=\prod{\cal L}_d$. 

The PDF ${ \cal P}_{\rm sig} (\dt,\, \sigdt; c)$, for each category  $c$, is the 
convolution of $F(\dt;\, c)$ (Eq.\ \ref{eq:FCPdef}) with the
signal resolution function (sum of three Gaussians) determined from the
\bflav\ sample.
The other PDF forms are: the sum of two Gaussians for ${\cal P}_{\rm
sig}(\mes)$ and ${\cal P}_{\rm sig}(\DE)$; the sum of three Gaussians for 
${\cal P}_{\qqbar}(\dt; c)$; a conjunction of two Gaussians with different
widths below and above the peak for ${\cal P}_j(\xf)$ (a small ``tail"
Gaussian is added for ${\cal P}_{\qqbar}(\xf)$); a linear
dependence for ${\cal P}_{\qqbar}(\DE)$; and for ${\cal
P}_{\qqbar}(\mes)$ the function
$x\sqrt{1-x^2}\exp{\left[-\xi(1-x^2)\right]}$, 
with $x\equiv2\mes/\sqrt{s}$.

For the signal and \BB\ background components we determine the PDF
parameters from simulation.  
For the \qqbar\ background we use
(\mes,\,\DE) sideband data to obtain initial values and ultimately leave them free to
vary in the final fit.

We compute the branching fractions and charge asymmetry from fits made
without \dt\ or flavor tagging, applied to candidates with \fetapreppgg\
and \fetaprg\ combined with \Kp\ or \kspppm.  
The free parameters in the fit are: the signal and \qqbar\ background yields, 
the peak position and lower and upper width
parameters of ${\cal P}_j(\xf)$ for signal and \qqbar\ background,
the tail fraction for ${\cal P}_{\qqbar}(\xf)$,
the slope of ${\cal P}_{\qqbar}(\DE)$ and $\xi$,
the width of the core Gaussian of ${\cal P}_{\rm sig}(\DE)$, the
mean of the core Gaussian of ${\cal P}_{\rm sig}(\mes)$, 
$n_{BB}$ for $B\ra\etapr_{\rho\gamma}K$, and for charged modes
the signal and background \acp. 

Table \ref{t:results} lists the quantities used to determine the branching 
fraction.  Equal production rates of \BpBm and \BzBzb  
pairs have been assumed.  To study biases from the likelihood fit, we
apply the method to simulated samples constructed to contain the 
signal and  background populations expected for data.  The resulting
yield biases, from unmodeled correlations in the signal PDF, are about
4\% for the measurements with \etaprrg, and negligible for those with
\fetapreppgg. The purity estimate in Table \ref{t:results} is given by the
ratio of the signal yield to the effective background plus signal, the
latter being defined as the square of the error on the yield.

\providecommand{\bfemsix}{${\cal B} (10^{-6})$}
\providecommand{\msp}{\phantom{-}}
\begin{table}[t]
\caption{
Signal yield, purity $P$(\%), reconstruction
efficiency $\epsilon$(\%), daughter branching fraction product,
measured branching fraction (\calB) in units of $10^{-6}$,
and \acp.}
\label{t:results}
\begin{tabular}{lcccccc}
\dbline
Mode		& Yield		& $P$   & $\epsilon$   &$\prod\calB_i$& \calB 	& $\acp\ (10^{-2})$ 	\\
\tbline 
$\fetapreppggk$                & $609\pm28$& 78& 23	& 0.175 & $66\pm3$&$-0.1\pm4.4$	\\
\fetaprgKp    & $1347\pm57$	& 41	& 26	& 0.295 & $72\pm3$&$\msp5.5\pm3.6$	\\
\fetapKp      &combined		&	&	&	& $69\pm2$&$\msp3.3\pm2.8$   \\
\tbline
$\fetapreppggkz_{\pi^+\pi^-}$	& $198\pm16$	& 77	& 23	& 0.060	& $61\pm5$&	\\
$\fetaprgKz_{\pi^+\pi^-}$	& $457\pm30$	& 51	& 26	& 0.102	& $73\pm5$&	\\
\fetapKz        &combined		&	&	&	& $68\pm3$&	\\
\dbline
\end{tabular}
\vspace{-5mm}
\end{table}

\begin{figure}[!htb]
 \includegraphics[angle=0,scale=0.4335]{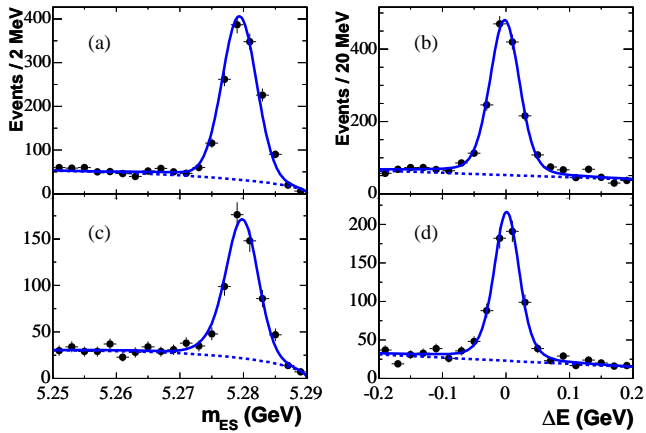}
\vspace{-0.7cm}
 \caption{\label{fig:projMbDE}
 The $B$ candidate \mb\ and \DE\ projections for \etapKp\ (a, b) and
\etapKzs\ (c, d).  Points with error bars represent the data, the solid
line the fit function, and the dashed line its background component.} 
\vspace{-.2cm}
\end{figure}

In Fig.\ \ref{fig:projMbDE}\ we show projections onto \mb\ and \DE\ for
a subset of the data for which the signal likelihood
(computed without the variable plotted) exceeds a mode-dependent
threshold that optimizes the sensitivity.

For the time-dependent analysis, we require $|\dt|<20~ps$ and $\sigdt<2.5~ps$.
We improve the sample size by combining the five decay chains listed in Table
\ref{t:resultsTD} 
in a single fit with 98 free parameters: $S$, $C$, signal yields (5),
$\etapr_{\rho\gamma} K^0$ \BB\ background yields (2), continuum background
yields (5) and fractions (30), background \dt,
\mes, \DE, \xf\ PDF parameters (54). The parameters $\tau$ and 
$\deltamd$ are fixed to world-average values \cite{PDG2004}.

\begin{figure}[!htb]
  \begin{center}
   \includegraphics[scale=0.3]{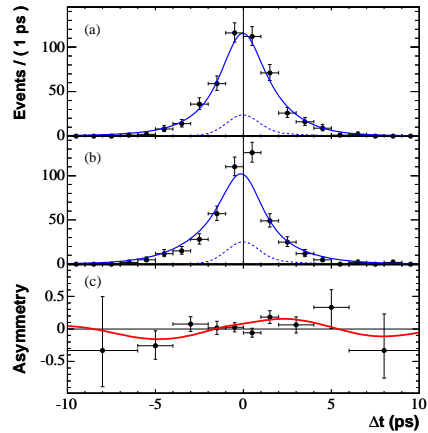}
\end{center}
  \vspace*{-0.5cm}
 \caption{Projections onto $\Delta t$ for \etapKzs\ of the data (points with error bars),
fit function (solid line), and background function
(dashed line), for (a) \Bz\ and (b) \Bzb\ tagged events, and (c) the
asymmetry between \Bz\ and \Bzb\ tags.}
  \label{fig:DeltaTProj}
\end{figure}

Table \ref{t:resultsTD} 
gives the yields, $S$ and $C$, and  
Fig.~\ref{fig:DeltaTProj} the $\Delta t$
projections and asymmetry of the combined neutral modes for 
events selected as for Fig.~\ref{fig:projMbDE}.

\begin{table}[!htb]
\caption{Results with statistical errors for the $\Bz\to\etapr\KS$
time-dependent fits.} 
\label{t:resultsTD}
\begin{center}
\vspace*{-0.3cm}
\begin{tabular}{lccc}
\dbline
Mode                     &Signal yield &       $S$       &         $C$      \\
\tbline
$\fetapreppggkz_{\pi^+\pi^-}$      & $188\pm15$  &$\msp0.01\pm0.28$&   $-0.18\pm0.18$ \\
$\fetaprgKz_{\pi^+\pi^-}$       & $430\pm26$  &$\msp0.44\pm0.19$&   $-0.30\pm0.13$ \\
$\fetapreppthrpikz_{\pi^+\pi^-}$&$54\pm8$ &$\msp0.79\pm0.47$&$\msp0.11\pm0.35$ \\
$\fetapreppggkz_{\pi^0\pi^0}$      &$44\pm9$ &   $-0.04\pm0.57$&   $-0.65\pm0.42$ \\
$\fetaprgKz_{\pi^0\pi^0}$       &  $94\pm23$  &   $-0.45\pm0.68$&$\msp0.41\pm0.40$ \\
\tbline
Combined fit                    & $804\pm40$  &$\msp0.30\pm0.14$&   $-0.21\pm0.10$ \\
\dbline
\end{tabular}
\end{center}
\vspace*{-0.3cm}
\end{table}

The major systematic uncertainties affecting  the branching fraction measurements 
reflect  the imperfect knowledge of the \etapr\ 
branching fractions (3.4\%) \cite{PDG2004}, and of the reconstruction efficiency 
 (0.8\%\ per charged track, 1.5\%\ per photon, and 2.1\%\ per \KS) estimated from 
auxiliary studies.  We take one-half of the
measured yield bias (0--2\%) as a systematic error.
Bias and systematic uncertainties for \acp\ have been estimated from
the values obtained for the background component in the fit to the
data.  We apply a correction of $+0.016$ and assign a systematic error of
$0.005$. 

For the time-dependent measurements, we find approximately equal (0.01) 
systematic uncertainties from several sources: variation of the signal  
PDF shape parameters within their errors, SVT alignment, position and size of the beam
spot, \BB\ background, modeling of the signal \dt\ distribution, and
interference between the CKM-suppressed $\bar{b}\to\bar{u} c\bar{d}$ amplitude 
and the favored $b\to c\bar{u}d$ amplitude for some tag-side $B$ decays
\cite{dcsd}. 
The \bflav\ sample is used  to determine the errors associated with the signal \dt\
resolutions, tagging efficiencies, and mistag rates, and published
measurements \cite{PDG2004} for $\tau_B$ and \deltamd.  
Summing all systematic errors in quadrature, we obtain 0.02 for $S$ and $C$.

In conclusion, we have used samples of about 2000 \etapKp\ and 800 \etapKzs\
events to measure  the branching fractions, the time-integrated charge 
asymmetry  and  the time-dependent asymmetry parameters \skz\ and \ckz.
The measured 
branching fractions are $\BretapKp=\RetapKp$ and $\BretapKz=\RetapKz$, and 
the charge asymmetry is $\acp = \rAetapKp$.  These precise branching fraction 
measurements challenge the theoretical understanding of these
decays  \cite{BrFr}. The measured charge asymmetry is consistent with zero,
with 90\% CL interval \RAetapKp, and constrains the amount of 
possible direct \CP\ violation in \BetapK\ decays.

The measured time-dependent \CP\ violation parameters in \BetapKzs\ are
 $\skz = \rSetapKz$ and $\ckz = \rCetapKz$.  Our result for \skz\ differs from that measured 
 by  \babar\  in $B^0\goto J/\psi\kzs$ \cite{s2b} by 3.0 standard deviations;
 it  also represents an improvement by a factor 2.4 (1.9) in precision
over the published results of \babar\ \cite{Previous} (Belle \cite{BELLE}).  
All these measurements supersede our previous published results \cite{Previous}.

We are grateful for the excellent luminosity and machine conditions
provided by our \pep2\ colleagues, 
and for the substantial dedicated effort from
the computing organizations that support \babar.
The collaborating institutions wish to thank 
SLAC for its support and kind hospitality. 
This work is supported by
DOE
and NSF (USA),
NSERC (Canada),
IHEP (China),
CEA and
CNRS-IN2P3
(France),
BMBF and DFG
(Germany),
INFN (Italy),
FOM (The Netherlands),
NFR (Norway),
MIST (Russia), and
PPARC (United Kingdom). 
Individuals have received support from CONACyT (Mexico), A.~P.~Sloan Foundation, 
Research Corporation,
and Alexander von Humboldt Foundation.


\begin{thebibliography}{99}
\bibitem{conjugates}
Charge conjugate decay modes are implied unless explicitly stated.
\end{thebibliography}
\end{document}